\documentclass[fleqn,usenatbib]{mnras}

\usepackage{newtxtext,newtxmath}

\usepackage[T1]{fontenc}

\DeclareRobustCommand{\VAN}[3]{#2}
\let\VANthebibliography\thebibliography
\def\thebibliography{\DeclareRobustCommand{\VAN}[3]{##3}\VANthebibliography}

\usepackage{graphicx}	
\usepackage{amsmath}	

\usepackage{subcaption}
\usepackage{algorithm}
\usepackage[noend]{algpseudocode}

\newcommand{\LineComment}[1]{\State $\triangleright$ \textit{#1}}

\title[Optimization of the Tree Algorithm]{Optimizing the Gravitational Tree Algorithm for Many-Core Processors}

\author[T. Tokuue and T. Ishiyama]{
Tomoyuki Tokuue$^{1}$\thanks{E-mail: t.tokuue@chiba-u.jp}
and Tomoaki Ishiyama$^{2}$\thanks{E-mail: ishiyama@chiba-u.jp}
\\
$^{1}$Department of Applied and Cognitive Informatics, Division of Mathematics and Informatics, Graduate School of Science and Engineering, Chiba University, 1-33, Yayoi-cho, Inage-ku, Chiba 263-8522, Japan\\
$^{2}$Digital Transformation Enhancement Council, Chiba University, 1-33, Yayoi-cho, Inage-ku, Chiba 263-8522, Japan
}

\date{Accepted XXX. Received YYY; in original form ZZZ}

\pubyear{2023}

\begin{document}
\label{firstpage}
\pagerange{\pageref{firstpage}--\pageref{lastpage}}
\maketitle

\begin{abstract}
Gravitational $N$-body simulations calculate numerous interactions between particles. The tree algorithm reduces these calculations by constructing a hierarchical oct-tree structure and approximating gravitational forces on particles. Over the last three decades, the tree algorithm has been extensively used in large-scale simulations, and its parallelization in distributed memory environments has been well studied. However, recent supercomputers are equipped with many CPU cores per node, and optimizations of the tree construction in shared memory environments are becoming crucial. We propose a novel tree construction method in contrast to the conventional top-down approach. It first creates all leaf cells without traversing the tree and then constructs the remaining cells by a bottom-up approach. We evaluated the performance of our novel method on the supercomputer Fugaku and an Intel machine. On a single thread, our method accelerates one of the most time-consuming processes of the conventional tree construction method by a factor of above 3.0 on Fugaku and 2.2 on the Intel machine. Furthermore, as the number of threads increases, our parallel tree construction time reduces considerably. Compared to the conventional sequential tree construction method, we achieve a speedup of over 45 on 48 threads of Fugaku and more than 56 on 112 threads of the Intel machine. In stark contrast to the conventional method, the tree construction with our method no longer constitutes a bottleneck in the tree algorithm, even when using many threads.
\end{abstract}




\section{Introduction}

$N$-body simulations numerically calculate particle-particle interactions to simulate the evolution of $N$-particle systems and have been used in various fields, including astrophysics and cosmology. Direct summation is the most basic algorithm for $N$-body simulations that directly calculates the gravitational forces between all particle pairs. It is the most accurate, but the time complexity is $O(N^2)$, making simulations with many particles impractical. To reduce the calculation cost, \citet{barnes_hierarchical_1986} proposed the tree algorithm that approximates the gravitational forces with the time complexity of $O(N\log N)$. It first constructs a hierarchical oct-tree structure based on the distribution of particles (``tree construction''). Each node of the tree represents a cubic region and is also referred to as "cell"~\footnote{We use the term "cell" to represent nodes of tree structures, as the term "node" is also used to represent units of computation.}. Then, it traverses the tree and approximately calculates gravitational forces on each particle from cells by evaluating their multipole expansions (``gravity calculation'').

Nowadays, the tree algorithm is regarded as a standard algorithm for gravitational $N$-body simulations and has been used for simulations with numerous particles. The number of particles in recent large-scale simulations sometimes exceeds a trillion~\citep[e.g.,][see also \citealt{Angulo2022} and references therein]{ishiyama_uchuu_2021,Wang2022,Frontiere2022}. To achieve such huge simulations, various approaches to parallelize the tree algorithm for distributed memory environments have been proposed over the last three decades~\citep[e.g.,][]{warren_parallel_1993,dubinski_parallel_1996,pangfeng_liu_experiences_2000,springel_gadget_2001,makino_fast_2004,warren_2hot_2013,iwasawa_implementation_2016}.
The Tree Particle-Mesh~\citep[TreePM; ][]{xu_new_1995} is a reincarnation of the tree algorithm coupled with the Particle-Mesh~\citep[PM; ][]{Hockney1981}, which is widely used in cosmological $N$-body simulations, and has also been parallelized for distributed memory environments~\citep[e.g.,][]{bagla_treepm_2002,Dubinski2004,Yoshikawa2005,springel_cosmological_2005,ishiyama_greem_2009,Ishiyama2012,Habib2012,Wang2018,springel_simulating_2021,Ishiyama2022}.

As the number of CPU cores in recent processors is significantly increasing, optimizations for shared memory environments are becoming increasingly important in addition to those for distributed memory environments. For instance, the computational node of the Japanese flagship supercomputer Fugaku at the RIKEN Center for Computational Science consists of one FUJITSU A64FX processor with 48 computational cores~\footnote{\url{https://www.r-ccs.riken.jp/en/fugaku/about/}}, which is six times more than its predecessor, the K computer's SPARC64 VIIIfx~\citep{yoshida2012sparc64}. This trend should continue because enhancing processor performance through increased operating frequency is approaching its limits in terms of power consumption. In modern massively parallel machines, shared-memory parallelization is favored within the computational node in addition to distributed-memory parallelization because accessing shared memory is faster than inter-process communication, enabling more efficient utilization of many cores. Multithreading also has the advantage of lower memory consumption than multiprocessing. One reason is that multithreading does not require replication of data shared among threads. Another reason is that inter-process communication requires each process to have its own communication buffer.

In shared-memory parallelization, the gravity calculation in the tree algorithm is expected to be sped up in proportion to roughly the number of threads because it can be performed independently for each particle or group of particles. \citet{hernquist_performance_1987} proposed creating an interaction list of cells before calculating the forces from them to parallelize the calculation efficiently. To reduce the tree traversal cost and enhance the parallel granularity, \citet{barnes_modified_1990} presented Barnes' modified algorithm that shares an interaction list with particles in a cell containing $n_\mathrm{crit}$ or fewer particles. Using those modifications, we can achieve extremely high efficiency for the gravity calculation in shared-memory parallelization. SIMD (Single Instruction, Multiple Data) instruction sets can further accelerate the calculation of gravitational interactions in proportion to the SIMD width~\citep{nitadori_performance_2006,tanikawa_n-body_2012,tanikawa_phantom-grape_2013,kodama_acceleration_2019}.

In contrast, the tree construction is more challenging to parallelize efficiently due to the dependency between hierarchical levels of the tree and considerable memory access. Unless the tree construction is highly parallelized, its time will occupy a more significant fraction of the total computational time with increasing threads. Additionally, it is difficult to accelerate the tree construction using SIMD instruction sets due to numerous nested conditional branches and memory access. Therefore, as the SIMD width increases, the proportion of the time for the tree construction increases.

For those reasons, a fast tree construction method is crucial to perform simulations efficiently and practically, especially on many-core processors with large SIMD registers. Despite its importance, there has been little effort to optimize the tree construction in shared memory environments, although there are optimized implementations for graphics processing units~\citep[e.g.,][]{bedorf_sparse_2012,bedorf_2477_2014,miki_gothic_2017,miki_gravitational_2019,keller_cornerstone_2023}. In this study, we propose a novel method to accelerate the tree construction on modern many-core processors.

This paper is organized as follows. In section~\ref{sec:methods}, we introduce the conventional tree algorithm and propose our optimized tree construction method and its parallelization. In section~\ref{sec:evaluation}, we evaluate the performance of our method using two distinct processors, FUJITSU A64FX and Intel Xeon Platinum 8280L. Finally, in section~\ref{sec:conclusions}, we summarize the results and discuss approaches to accelerate the tree construction further.

\section{Tree algorithm and our approaches}\label{sec:methods}

In section~\ref{sec:bhtree}, we briefly review the original tree algorithm presented by \citet{barnes_hierarchical_1986}, which calculates the gravitational forces on $N$ particles in $O(N\log N)$ time complexity. We then introduce an efficient tree construction method presented by \citet{warren_parallel_1993} in section~\ref{sec:hotree} and describe our optimized tree construction method and its parallelization strategies in section~\ref{sec:our_method}.

\subsection{Barnes--Hut tree}\label{sec:bhtree}
The tree algorithm proposed by \citet{barnes_hierarchical_1986} consists of ``tree construction'' and ``gravity calculation''. The tree construction starts with a root cubic cell that contains all particles, and the root cell is recursively halved along each dimension until every cell holds at most $n_\mathrm{leaf}$ particles. During division, particles in a cell are distributed among its newly created eight child cells. Once the oct-tree structure is constructed, the algorithm traverses the tree from the root to calculate the gravitational forces on each particle from cells. The gravitational force on a particle is approximated by evaluating the cell's multipole expansions if the following condition is satisfied,
\begin{equation}
\label{eq:condition_approximate_cell}
\frac{l}{d}<\theta,
\end{equation}
where $d$ is the distance between the particle and the cell's center of mass, and $l$ is the cell's side length. The constant $\theta$ is a free parameter that determines the accuracy of the approximation. A larger value of $\theta$ results in a higher error and a shorter computation time. If a cell does not satisfy the condition, the gravitational force is evaluated by recursively accumulating forces from its eight child cells.
Barnes' modified algorithm can further accelerate the gravity calculation~\citep{barnes_modified_1990} by letting particles in a cell containing $n_\mathrm{crit}$ or fewer particles share their interaction list.

\subsection{Hashed oct-tree}\label{sec:hotree}
\citet{warren_parallel_1993} proposed a novel data structure for the tree algorithm, the hashed oct-tree. It assigns a unique integer key to each cell and allows access to any cell using a hash table instead of traversing the tree with standard pointers. The cell keys are generated based on the Morton key~\footnote{The Peano--Hilbert ordering, sometimes used for domain decomposition, can replace the Morton ordering.}~\citep{morton1966computer}, with the center of the root cell as the origin. The root cell is represented by the key of 1. The key for any other cell is generated by concatenating its parent cell's key in binary representation and a 3-bit integer, which indicates whether the cell is located in the left or right half of its parent cell along the x-, y-, and z-axes.

They also introduced an efficient tree construction method that utilizes this tree data structure. The method starts by calculating the 64-bit Morton keys for particles based on their positions. First, each floating point coordinate is converted into a 21-bit integer for every dimension. Then, these integers are interleaved, with the most significant bit of 1. By forcing the side length of the root cell to a power of two and setting the origin of the Morton key at its center, we can associate the particle keys with the cell keys. Namely, the upper bits of a particle key hold the keys of cells containing the particle.

This method sorts particles by their keys and traverses the tree to assign each particle to an appropriate leaf while creating intermediate cells. It begins traversing the tree from the cell to which the previous particle was assigned instead of always beginning at the root cell. Since particles with similar keys are located closely within the tree, this approach reduces the time to traverse the tree and find the leaf encompassing a given particle.

\subsection{Our optimized tree construction method}\label{sec:our_method}
Although the algorithm presented by \citet{warren_parallel_1993} is ground-breaking, it may not be optimal for the parallelization on many-core shared-memory environments because such machines were not widespread when they proposed their algorithm.

In this paper, we advance the hashed oct-tree's optimization and introduce a novel tree construction method that does not re-assign particles in the tree data structure. The overall procedure consists of the following four steps:
\begin{enumerate}
\item Initializing a hash table and calculating particle keys
\item Sorting particle keys paired with their indices
\item Reordering particles according to the sorted keys
\item Creating tree cells beginning with leaf cells by a bottom-up approach
\end{enumerate}
The sorting procedure is done in two steps to minimize memory copying. In practice, the reordering of particles is optional and not necessarily required at every simulation step, but it enhances cache efficiency when accessing particles. We perform the reordering at every step for our performance evaluation. After the sorting, we create cells using a new approach that initially generates all leaf cells based on the sorted particle keys and subsequently builds non-leaf cells by a bottom-up approach.
Another advantage of this method is that it can calculate the properties of cells (e.g., multipole expansions) as soon as they are created without recursive traversing of the tree after the construction.

\subsubsection{Creating leaves from sorted particle keys}

We define the depth of a cell as the number of edges from the root cell in the tree, e.g., the root cell's depth is 0, and the depth of its child cells is 1. Since we use the identical key generation algorithm for particles and cells, the key of a cell at depth $d$ matches the upper $3d+1$ bits of the keys of particles in that cell. In other words, we can derive a leaf's key from its depth and the key of a particle in the leaf without traversing the tree.

We determine the depth of the leaf containing a given particle based on its key. Here, we define ``the minimum distinguished depth'' as the minimum depth at which two distinct particles are assigned to different cells. For example, a particle with the key $1010011\dots$ and one with the key $1011011\dots$ are in the same cell 1 (the root) at depth 0 but in different cells 1010 and 1011 at depth 1, respectively. Thus, the minimum distinguished depth of them is 1.

Sorting the particle keys arranges the keys of the particles in a cell contiguously because they share the same upper bits. We denote the particle with the $k$-th key in the sorted order as "particle $k$" and "the minimum distinguished depth of particle $i$ and particle $j$" as $\text{minDepth}(i,j)$. If particle $k$ is the first particle in a leaf, the leaf depth $D_k$ matches the larger of $\text{minDepth}(k,k-1)$ and $\text{minDepth}(k,k+n_\mathrm{leaf})$ (see Appendix~\ref{sec:proof_of_the_formula} for the proof). We extract the upper $3D_k+1$ bits from particle $k$'s key to determine its leaf key. We then identify all the particles in this leaf by sequentially scanning the sorted particle keys from the $k$-th key until the upper bits do not match the leaf key. Algorithm~\ref{code:create_leaves} shows the pseudocode to create all leaf cells from sorted particle keys. To calculate minDepth$(i,j)$, we first count the number of leading zero bits in the XOR (exclusive or) of particle $i$'s and particle $j$'s keys, which is the number of leading bits shared by those keys and can be counted using various methods, e.g., the GNU Compiler Collection (GCC) built-in function \_\_builtin\_clzl. Then, we divide this number by three and round up to obtain minDepth$(i,j)$.

\begin{algorithm}
    \caption{Creating leaves from sorted $N$ particle keys.}
    \label{code:create_leaves}
    \begin{algorithmic}[1]
        \Procedure{CreateLeaves}{}
        \State $k\gets 0$
        \While{$k<N$} \Comment{Particle $k$ is the first particle in a leaf}
        \State $D_k\gets 0$ \Comment{$D_k$ is the leaf depth}
        \If{$k>0$}
        \State $D_k\gets\Call{MinDepth}{k-1,k}$
        \EndIf
        \If{$k+n_\mathrm{leaf}<N$}
        \State $D_k\gets\max\{D_k,\Call{MinDepth}{k,k+n_\mathrm{leaf}}\}$
        \EndIf
        \State leafKey $\gets$ \Call{ExtractCellKey}{$k,D_k$}
        \LineComment{Create a new leaf cell}
        \State leaf $\gets$ \Call{CreateLeaf}{leafKey}
        \While{$k<N$} \Comment{Check if particle $k$ is in the leaf}
        \If{\Call{ExtractCellKey}{$k,D_k$} $=$ leafKey}
        \LineComment{Assign particle $k$ to the leaf}
        \State \Call{AssignParticle}{$k$, leaf}
        \State $k\gets k + 1$
        \Else
        \LineComment{All the particles have been assigned to the leaf}
        \LineComment{Particle $k$ is the first particle in the next leaf}
        \State \textbf{break}
        \EndIf
        \EndWhile
        \EndWhile
        \EndProcedure
        \State
        \Procedure{MinDepth}{$i,j$}
        \State xor $\gets$ \Call{BitwiseXOR}{particle $i$'s key, particle $j$'s key}
        \State \Return $\lceil{\Call{CountLeadingZeroBits}{\text{xor}}/3}\rceil$
        \EndProcedure
        \State
        \Procedure{ExtractCellKey}{$k,d$}
        \LineComment{The upper $3d+1$ bits of a 64-bit key}
        \State \Return \Call{RightShift}{particle $k$'s key, $64-(3d+1)$}
        \EndProcedure
    \end{algorithmic}
\end{algorithm}

\subsubsection{Creating non-leaf cells}

After creating all the leaves, we create non-leaf cells by a bottom-up approach. We prepare a list of the keys~\footnote{In practice, storing a pointer to each created cell with its key in the list allows us to access child cells without referring to the hash table when calculating the center of mass of a cell.} of created cells for each depth and generate all non-leaf cells in descending order of depth by creating the parent cells of cells at depth $d$ and appending their keys to the list for depth $d-1$. We must avoid creating duplicate cells. Although binary search trees or hash tables can eliminate duplicates, we introduce a straightforward method that does not require additional data structures.

We separate the created cell keys into those of leaves and those of non-leaf cells. The leaf keys at each depth are sorted because we append them to the list in the order of sorted particle keys, taking the same number of leading bits from each key. If the non-leaf cell keys at depth $d$ are sorted, the merge operation in merge sort to combine them with the leaf keys at the same depth yields the sorted list of all cell keys at depth $d$. Then, we scan the merged list and register~\footnote{If a hash table represents the tree, it is sufficient to gather information on registered child cells, such as their number of particles. Otherwise, in addition to that, cells are linked with pointers.} each cell as a child of its parent cell. If the parent cell's key does not match the last created parent cell's key, the parent cell has not been created yet, and all the child cells of the last created parent cell have already been registered. Thus, we calculate the center of mass~\footnote{Other properties, such as higher-order multipoles, can be calculated similarly. We calculate the geometric center of a cell using the level and key of the cell when it becomes necessary.} of the last created parent cell from that of its child cells and create a new parent cell for the current cell. As a result, sorted non-leaf cell keys at depth $d-1$ are generated from those at depth $d$~\footnote{This procedure is akin to creating leaves from sorted particle keys.}. In the first place, the list of non-leaf cell keys at the maximum depth of leaves is empty, and the list of leaf keys at the same depth is sorted. Therefore, repeating this process in descending order of depth can produce all non-leaf cells. In reality, we can omit the merging of two sorted lists at each depth by iteratively comparing the first cell keys from the two sorted lists and creating a parent cell of the one with the smaller key (algorithm~\ref{code:create_parents}).

\begin{algorithm}
    \caption{Creating the parents of cells at depth $d$.}
    \label{code:create_parents}
    \begin{algorithmic}[1]
        \Require{$N_\mathrm{leaf}$ leaf keys and $N_\mathrm{inner}$ non-leaf cell keys at depth $d$}
        \Procedure{CreateParents}{leafKeys, innerCellKeys}
        \State $i,j\gets 0$ \Comment{Heads of the two sorted lists}
        \While{$i<N_\mathrm{leaf}$ and $j<N_\mathrm{inner}$}
        \LineComment{Construct the parent with the smaller key}
        \If{leafKeys$[i]$ $<$ innerCellKeys$[j]$}
        \State \Call{ConstructParent}{leafKeys$[i]$}
        \State $i\gets i + 1$
        \Else
        \State \Call{ConstructParent}{innerCellKeys$[j]$}
        \State $j\gets j + 1$
        \EndIf
        \EndWhile
        \LineComment{Scan through the remnants}
        \While{$i<N_\mathrm{leaf}$}
        \State \Call{ConstructParent}{leafKeys$[i]$}
        \State $i\gets i + 1$
        \EndWhile
        \While{$j<N_\mathrm{inner}$}
        \State \Call{ConstructParent}{innerCellKeys$[j]$}
        \State $j\gets j + 1$
        \EndWhile
        \EndProcedure
        \State
        \Procedure{ConstructParent}{cellKey}
        \State parentKey $\gets$ \Call{RightShift}{cellKey, 3}
        \If{parentKey $\neq$ lastParentKey}
        \LineComment{The construction of the last parent is complete}
        \LineComment{Create a new parent}
        \State lastParent $\gets$ \Call{CreateCell}{parentKey}
        \State lastParentKey $\gets$ parentKey
        \EndIf
        \LineComment{Register the cell as a child of the last created parent}
        \State \Call{AppendChild}{cellKey, lastParent}
        \EndProcedure
    \end{algorithmic}
\end{algorithm}

\subsubsection{Parallelization\protect\footnote{Please note that the parallelization strategies described in this section are not specific to our novel tree construction method. They should also be applied to conventional tree construction methods.}}
The initialization of the tree and the reordering of particles are easily parallelized by evenly distributing data among threads. We can parallelize the sorting of particle keys using known multi-threaded sorting algorithms. To parallelize the creation of cells, we first divide the tree into subtrees. We recursively divide the sorted particle keys into blocks until each block forms a subtree containing $n_\mathrm{serial}$ or fewer particles. This division is done by scanning the upper bits of particle keys, ensuring that different subtrees do not have the same cell. We then dynamically assign these blocks to threads as they can build subtrees in parallel. Each thread creates a subtree in the same way as in the non-parallel case.
However, since the depth of the root cell $d_\mathrm{root}$ of the subtree is non-zero, the formula for calculating the leaf depth $D_k$ changes:
\begin{equation}
D_k=\max\{\text{minDepth}(k,k-1),\text{minDepth}(k,k+n_\mathrm{leaf}),d_\mathrm{root}\}.
\end{equation}
After each thread finalizes the subtree construction, a thread constructs the remaining upper cells sequentially.

\citet{warren_parallel_1993} treat certain upper bits of a cell key as hash and associate the keys and the cells using a chained hash table. We record the mappings of the cells close to the root, which are frequently accessed, with an array to speed up the access to those cells and use a chained hash table for the other cells to reduce memory usage. The chained hash table needs to be updated in a thread-safe manner because cells created in parallel can have an identical hash. We ensure thread safety by implementing a non-blocking linked list~\citep{harris_pragmatic_2001} with the compare-and-swap instruction.

\section{Performance Evaluation}\label{sec:evaluation}
We conducted performance tests of our tree implementation on two computers with different architectures described in Table~\ref{tab:system_specs}. The first one is a single computational node of the supercomputer Fugaku at the RIKEN Center for Computational Science, equipped with one FUJITSU A64FX processor. This processor has four Core Memory Groups (CMGs~\footnote{Equivalent to NUMA nodes}), and each CMG has 12 computational cores. The second one is a large memory node provided by the Fugaku for pre/post analyses, comprising four Intel Xeon Platinum 8280L processors, each with 28 physical cores.
\begin{table*}
	\centering
    \caption{Specifications of the two systems used in the performance tests.}
	\label{tab:system_specs}
	\begin{tabular}{lccccccccc}
		\hline
        Name & Processor & Architecture & NUMA nodes & Cores & Frequency & Bandwidth & Compiler & Optimization options\\
        & & & & & GHz & GB/s & & \\
		\hline
        Fugaku & FUJITSU A64FX & aarch64 & 4 & $48\times1$ & 2.00 & 1024 & FCC (clang) & -Ofast \\
        Intel & Xeon Platinum 8280L & x86\_64 & 8 & $28\times4 $& 2.70 & 563 & GCC & -Ofast -march=native\\
		\hline
	\end{tabular}
\end{table*}

We simulated the time evolution of $N$-body systems using the leapfrog integrator over 320 constant timesteps and evaluated our method's performance based on the average elapsed time per step. To investigate the effect of particle distribution on the performance, we used two distinct particle distributions: Plummer sphere and exponential disk. We generated these distributions using NEMO package~\footnote{\url{https://github.com/teuben/nemo}}~\citep{teuben_stellar_1995}, commands \texttt{mkplummer} and \texttt{mkexpdisk}. The Plummer sphere has a higher density near its center and is dynamically stable. We also used the exponential disk to evaluate the performance of a non-spherically symmetric particle distribution. Unlike the Plummer, particles of the exponential disk are distributed in a thin disk shape.

We compared two implementations for creating tree cells: the conventional method~\citep{warren_parallel_1993} and ours presented in the previous section. The program was written in C++ and parallelized using the OpenMP library. We used \texttt{\#pragma omp for} directive and enabled dynamic scheduling only for the creation of cells and gravity calculation.
For sortings, we used a multi-threaded sorting code~\citep{tokuue_performance_2023}, which implements Parallel Sorting by Regular Sampling~\citep{shi_parallel_1992}, and BlockQuicksort~\citep{edelkamp_blockquicksort_2019} for sequential sorting.
We generated a highly optimized code by SIMD instruction sets for particle-particle gravitational interactions using the kernel generator PIKG~\footnote{\url{https://github.com/FDPS/PIKG}}. We set $\theta=0.75$, $n_\mathrm{leaf}=10$, and $n_\mathrm{crit}=512$ for the tree algorithm. Floating-point numbers were represented in double-precision, except in interaction calculations, where we used single-precision. We ensured that the forces were invariable regardless of the number of threads. For the parallelization of the tree construction, too small $n_\mathrm{serial}$ results in higher scheduling overhead and too large $n_\mathrm{serial}$ causes load imbalance. We find a power of two that minimizes the elapsed time on 48 threads and finally set $n_\mathrm{serial} = 32,768$.
In multiple non-uniform memory access (NUMA) systems, we pre-distribute particles and other resources across each NUMA node using \texttt{\#pragma omp for} directive and ``first-touch'' policy to improve memory access locality.

\subsection{Single-threaded performance}\label{sec:single_thread_performance}
\begin{figure*}
    \begin{subfigure}{0.8\linewidth}
        \includegraphics[width=\linewidth]{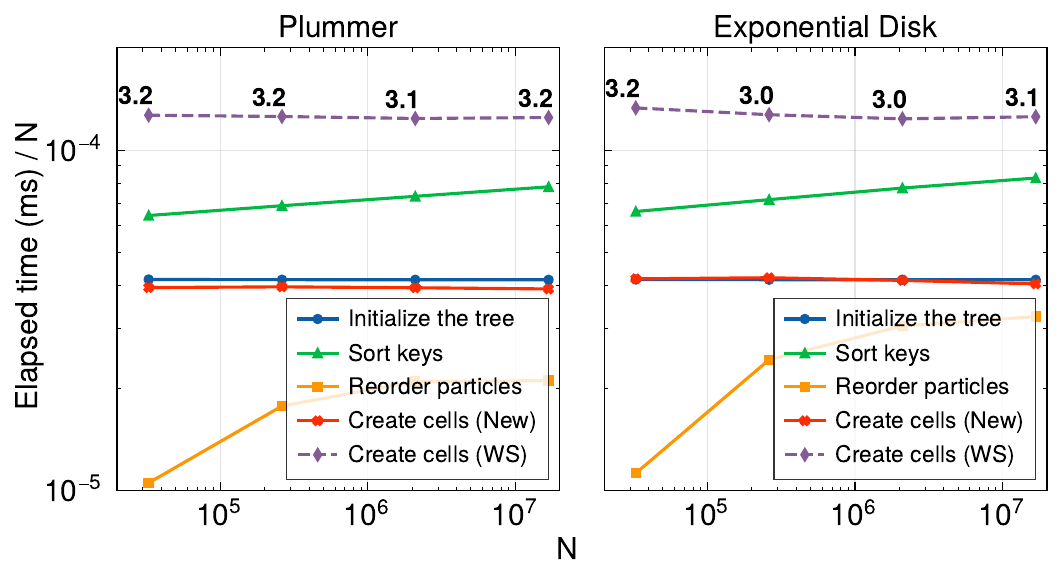}
        \caption{The Fugaku results.}
        \label{fig:serial_fugaku}
    \end{subfigure}
    \begin{subfigure}{0.8\linewidth}
        \includegraphics[width=\linewidth]{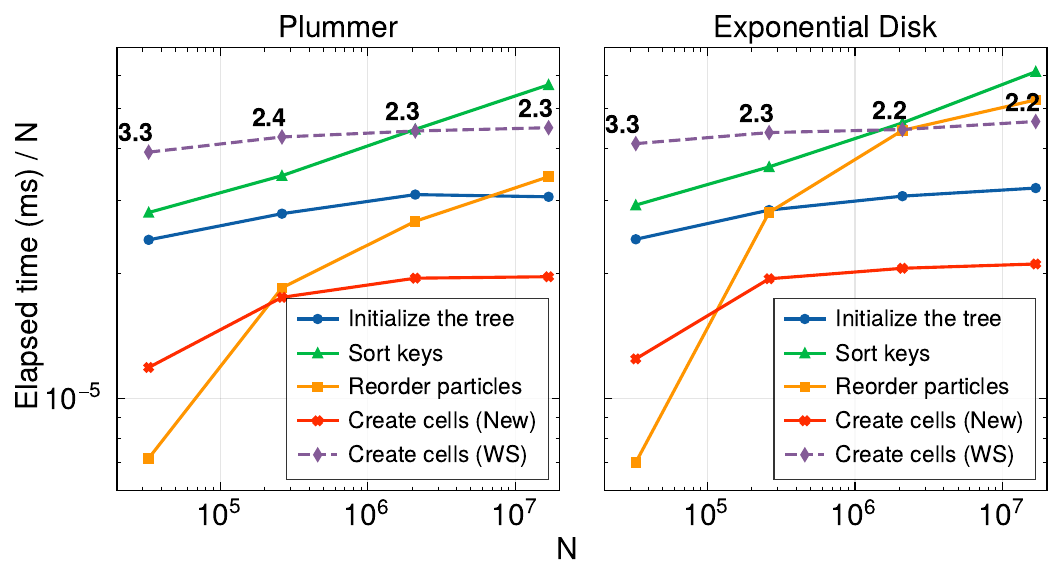}
        \caption{The Intel machine results.}
        \label{fig:serial_prepost}
    \end{subfigure}
    \caption{Breakdown of single-threaded performance in the tree construction. The horizontal axis represents the number of particles $N$ ranging from 32,768 to 16,777,216, while the vertical axis represents the average elapsed time divided by $N$. We compare the conventional method~\citep[WS: ][]{warren_parallel_1993} and our method (New) to create cells. Speedup factors of our method over the conventional method (time ratio of WS to New) are annotated along with the WS results.}
    \label{fig:serial}
\end{figure*}

We first investigate the performance of our code on a single thread and its dependency on the number of particles ($N=32,768$ to 16,777,216) and the initial particle distribution. We also compare the results with the conventional method~\citep{warren_parallel_1993}. Fig.~\ref{fig:serial} shows the average elapsed time for the four components of the tree construction divided by the number of particles.
The time taken for sorting keys gradually increases with $N$ due to its $O(N\log N)$ time complexity. In contrast, the time taken for initializing the tree, reordering particles, and creating cells are asymptotically constant with $N$, reflecting their $O(N)$ time complexity. They perform faster when $N$ is small because of a high cache hit ratio. This behavior is the most prominent for the reordering of particles because its performance is limited by memory access. It is more pronounced on the Intel machine, which has lower memory performance than Fugaku. Except for the reordering of particles, those results are more or less independent of the initial particle distributions.

Fig.~\ref{fig:serial_fugaku} shows that the creation of cells is the most time-consuming process of the conventional tree construction method on Fugaku, and we have succeeded in accelerating it by a factor of at least 3.0. Fig.~\ref{fig:serial_prepost} shows that even though the creation of cells occupies a smaller percentage on the Intel machine, we have achieved a speedup by a factor of at least 2.2. The time taken for creating cells in our novel method occupies about 21\,\% and 13\,\% of the total with $N=16,777,216$ on Fugaku and the Intel machine, respectively, suggesting that the creation of cells itself is no longer a bottleneck of the tree construction on a single thread, in stark contrast to the conventional method. However, those values also imply that it could reverse without achieving the high parallel efficiency for creating cells.

\subsection{Multi-threaded performance}
Next, we evaluate our parallel tree code's performance by varying the number of threads. In NUMA (and also CMG) systems, the memory location relative to processors influences overall performance, even using the same number of cores. To simplify the performance evaluation, we vary the number of cores in two distinct ways. First, we use only cores in a single NUMA node in which all the cores have uniform memory access costs. Since the total memory bandwidth remains constant within a single NUMA node, the memory bandwidth per core decreases as the number of cores increases. Secondly, we use multiple NUMA nodes to increase the number of cores further. In this test, we adopt a fixed number of cores per NUMA node, ensuring constant memory bandwidth per core regardless of the number of NUMA nodes used. In both cases, we bind one thread to one physical core and set the number of particles to 16,777,216.

\subsubsection{A single NUMA node}
\begin{figure*}
    \begin{subfigure}{0.8\linewidth}
        \includegraphics[width=\linewidth]{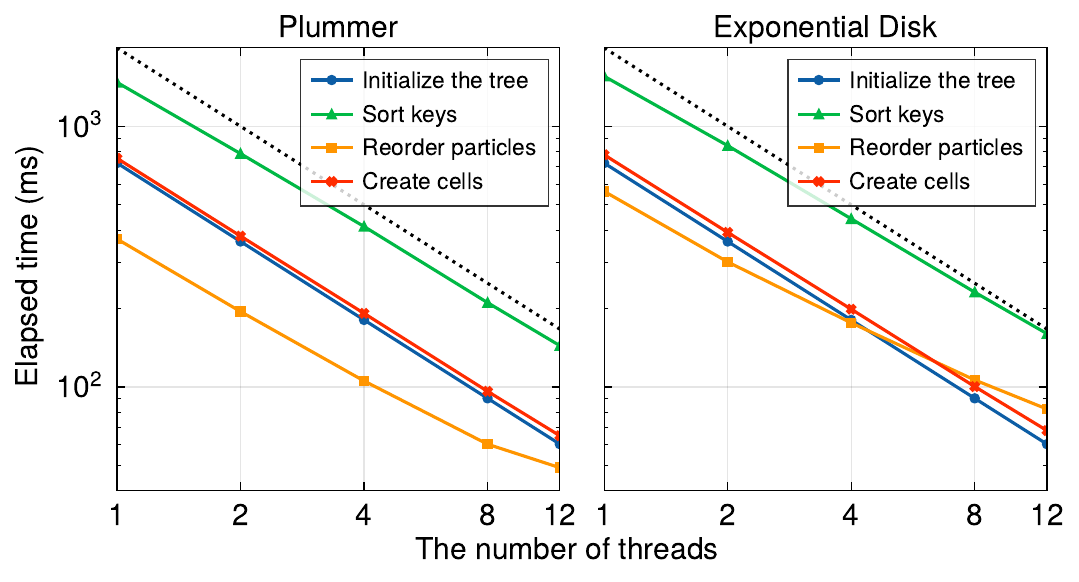}
        \caption{The elapsed time. The black dotted line shows an ideal scaling.}
        \label{fig:fugaku_tree_time1}
    \end{subfigure}
    \begin{subfigure}{0.8\linewidth}
        \includegraphics[width=\linewidth]{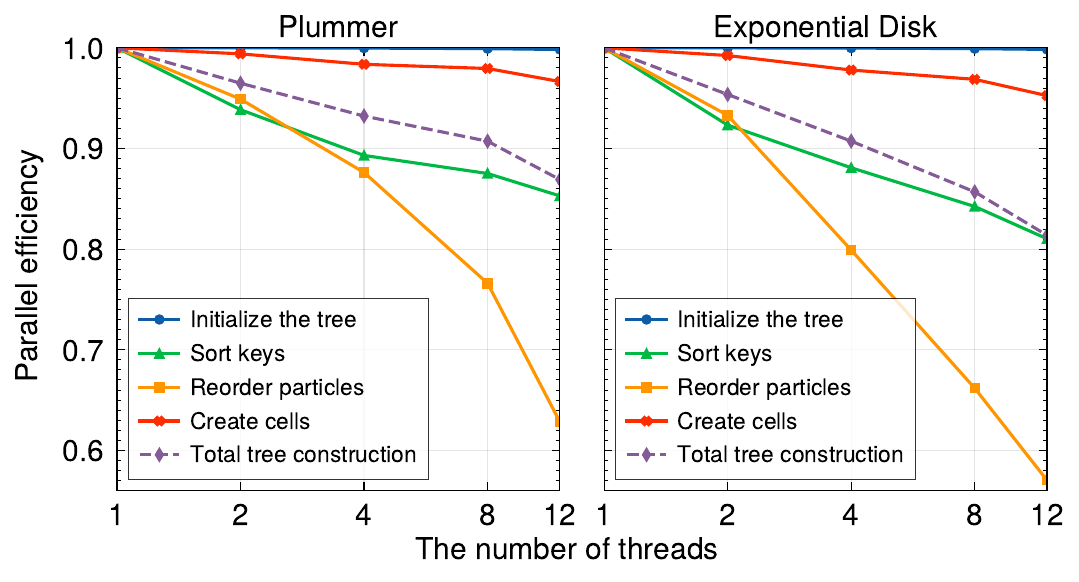}
        \caption{The parallel efficiency.}
        \label{fig:fugaku_tree_eff1}
    \end{subfigure}
    \caption{Breakdown of multi-threaded performance in the tree construction on Fugaku using a single CMG. The number of particles is fixed at 16,777,216.}
    \label{fig:fugaku_parallel1}
\end{figure*}
\begin{figure*}
    \begin{subfigure}{0.8\linewidth}
        \includegraphics[width=\linewidth]{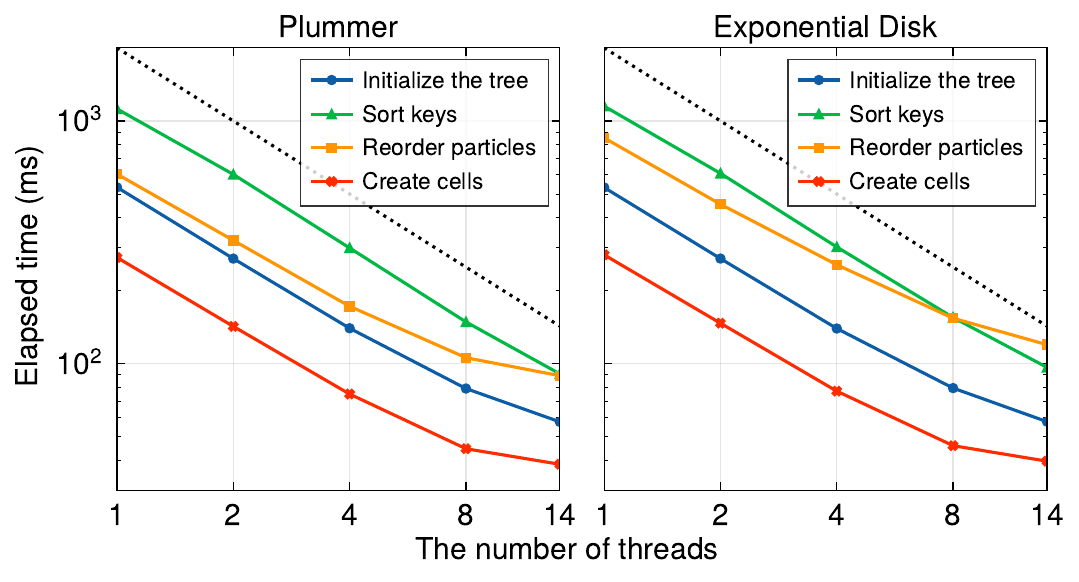}
        \caption{The elapsed time. The black dotted line shows an ideal scaling.}
        \label{fig:prepost_tree_time1}
    \end{subfigure}
    \begin{subfigure}{0.8\linewidth}
        \includegraphics[width=\linewidth]{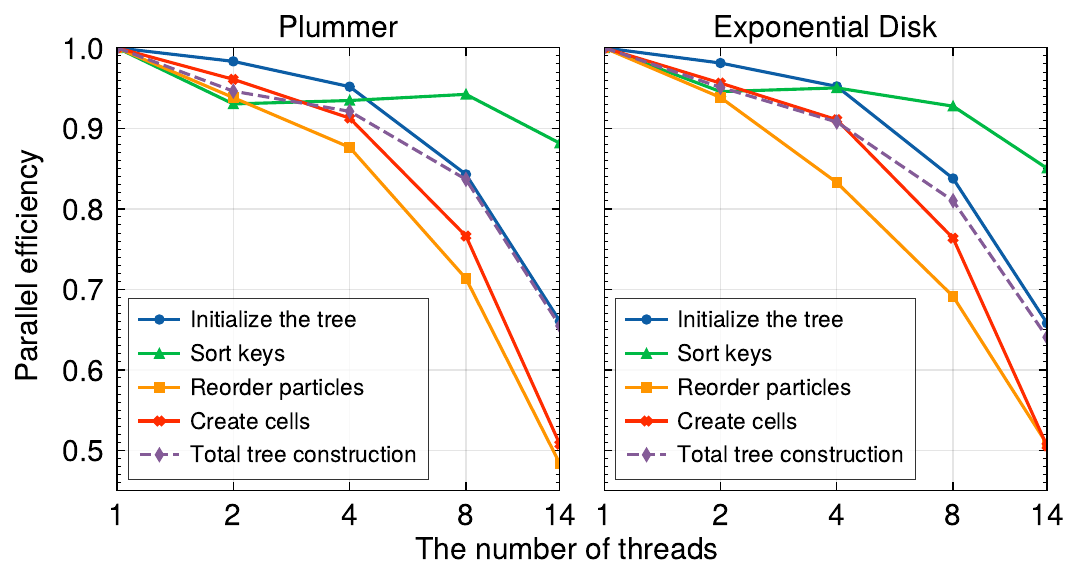}
        \caption{The parallel efficiency.}
        \label{fig:prepost_tree_eff1}
    \end{subfigure}
    \caption{Breakdown of multi-threaded performance in the tree construction on the Intel machine using a single NUMA node. The number of particles is fixed at 16,777,216.}
    \label{fig:prepost_parallel1}
\end{figure*}

Fig.~\ref{fig:fugaku_tree_time1} and~\ref{fig:fugaku_tree_eff1} show the average elapsed time for all the tree construction components and their parallel efficiency on a single CMG of Fugaku as a function of the number of threads, $p$. The parallel efficiency $e(p)$ is defined as
\begin{equation}
e(p) = \frac{t(1)}{t(p)}\frac{1}{p},
\label{formula:efficiency}
\end{equation}
where $t(p)$ is the average elapsed time with $p$ threads. The parallel efficiency indicates the degree of parallel performance compared to the ideal scaling, namely $t(p) \propto p^{-1}$.

The average elapsed time for all the tree construction components decreases monotonically, and the sorting of particle keys takes the most time. The time spent on reordering particles is the least when using fewer threads, but it becomes significant as the number of threads increases due to low parallel efficiency. The initialization of the tree and the creation of cells scale linearly with up to 12 threads regardless of the particle distributions. The parallel efficiency of the sorting of particle keys and the reordering of particles decreases as the number of threads increases. These processes involve large amounts of memory access, and their efficiency is strongly affected by reduced memory bandwidth with increasing the number of threads. In particular, most of the time for reordering particles is spent on memory copying, leading to the lowest parallel efficiency. The overall parallel efficiency of tree construction is mainly affected by the sorting of particle keys, but it keeps 81\,\% with up to 12 threads.

Fig.~\ref{fig:prepost_tree_time1} and~\ref{fig:prepost_tree_eff1} show the average elapsed time for all the tree construction components and their parallel efficiency on a single NUMA node of the Intel machine as a function of the number of threads. Fig.~\ref{fig:prepost_tree_time1} indicates that the creation of cells constitutes the least percentage in any case, and compared to Fugaku, the initialization of the tree and the reordering of particles account for a larger percentage regardless of the particle distribution. The time taken for reordering particles with 14 threads becomes comparable to or larger than that for sorting keys. Fig.~\ref{fig:prepost_tree_eff1} shows that, besides the reordering of particles, the initialization of the tree and the creation of cells suffer from sharp drops in efficiency, showing the stark difference from Fugaku. Since these processes show the ideal scaling on Fugaku, these drops are likely due to the Intel machine's lower memory bandwidth. Nevertheless, the overall parallel efficiency of the tree construction is above 63\,\% with up to 14 threads, reinforcing the effectiveness of our novel tree construction method.

\subsubsection{Multiple NUMA nodes}
\begin{figure*}
    \begin{subfigure}{0.8\linewidth}
        \includegraphics[width=\linewidth]{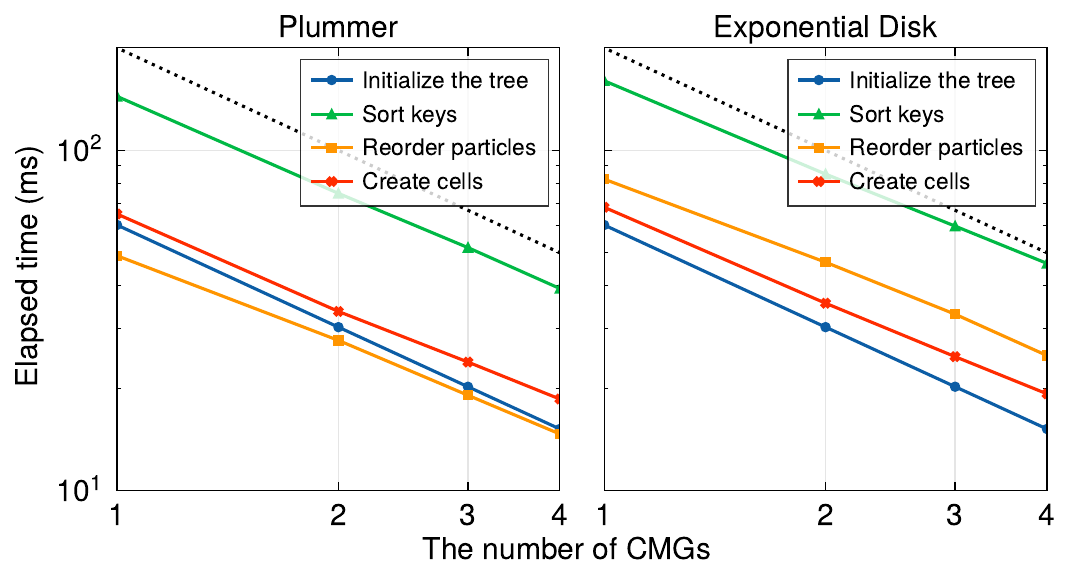}
        \caption{The elapsed time. The black dotted line shows an ideal scaling.}
        \label{fig:fugaku_tree_time2}
    \end{subfigure}
    \begin{subfigure}{0.8\linewidth}
        \includegraphics[width=\linewidth]{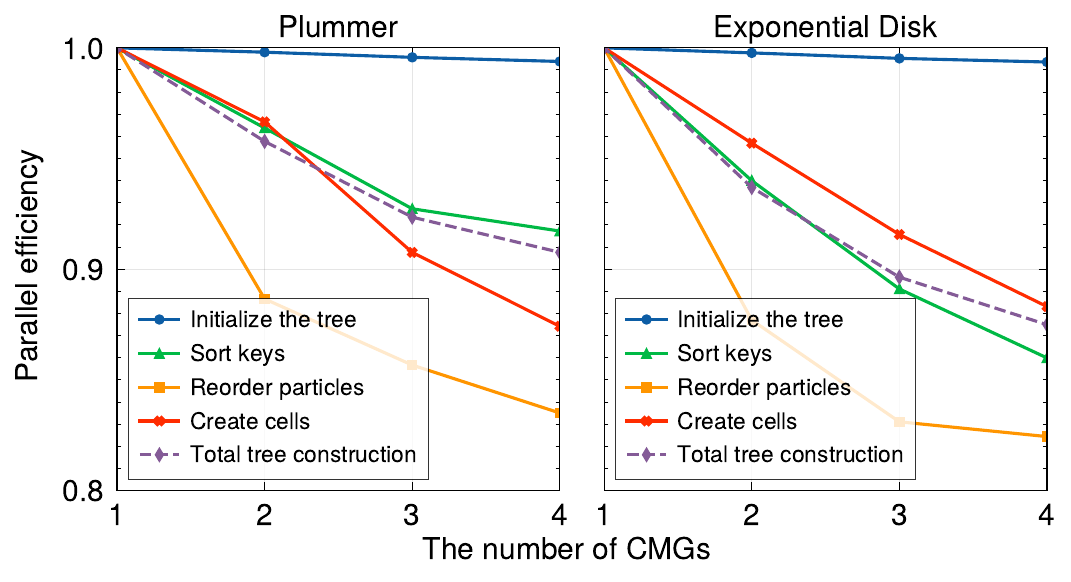}
        \caption{The parallel efficiency.}
        \label{fig:fugaku_tree_eff2}
    \end{subfigure}
    \caption{Breakdown of multi-threaded performance in the tree construction on Fugaku using multiple CMGs. The number of particles is fixed at 16,777,216.}
    \label{fig:fugaku_parallel2}
\end{figure*}
\begin{figure*}
    \begin{subfigure}{0.8\linewidth}
        \includegraphics[width=\linewidth]{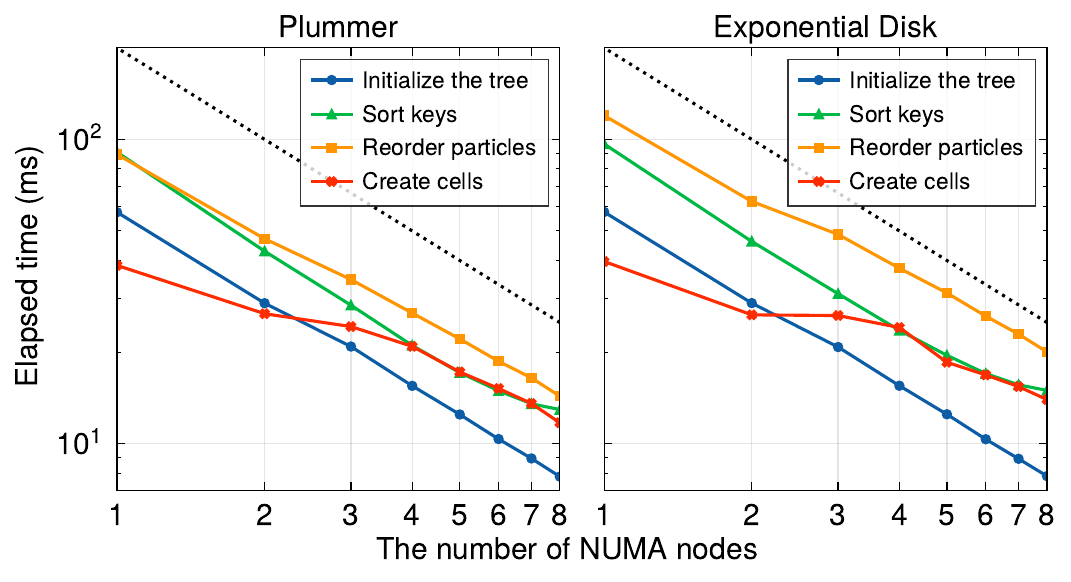}
        \caption{The elapsed time. The black dotted line shows an ideal scaling.}
        \label{fig:prepost_tree_time2}
    \end{subfigure}
    \begin{subfigure}{0.8\linewidth}
        \includegraphics[width=\linewidth]{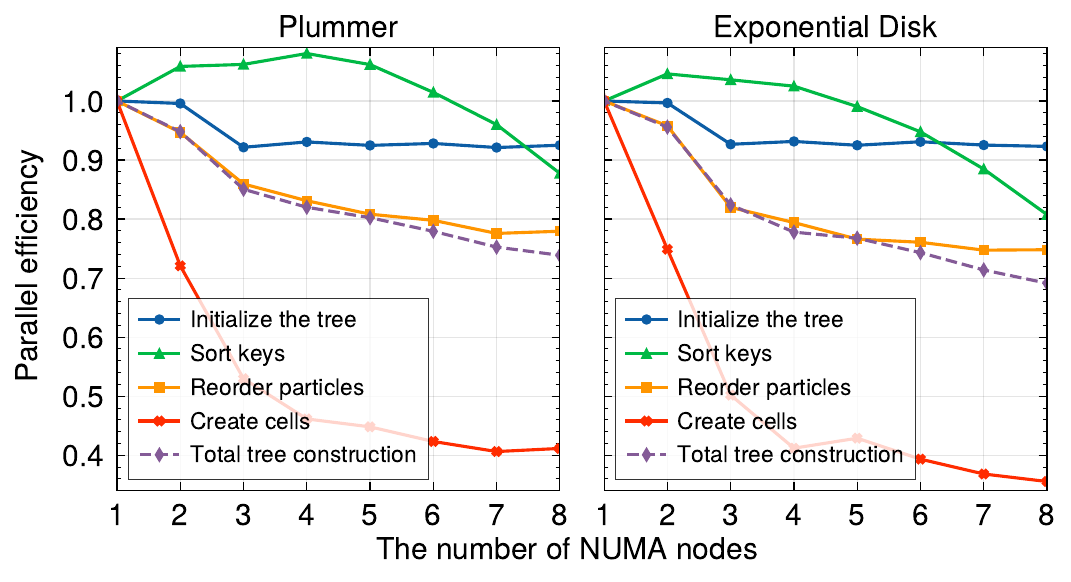}
        \caption{The parallel efficiency.}
        \label{fig:prepost_tree_eff2}
    \end{subfigure}
    \caption{Breakdown of multi-threaded performance in the tree construction on the Intel machine using multiple NUMA nodes. The number of particles is fixed at 16,777,216.}
    \label{fig:prepost_parallel2}
\end{figure*}
\begin{figure*}
    \begin{subfigure}{0.8\linewidth}
        \includegraphics[width=\linewidth]{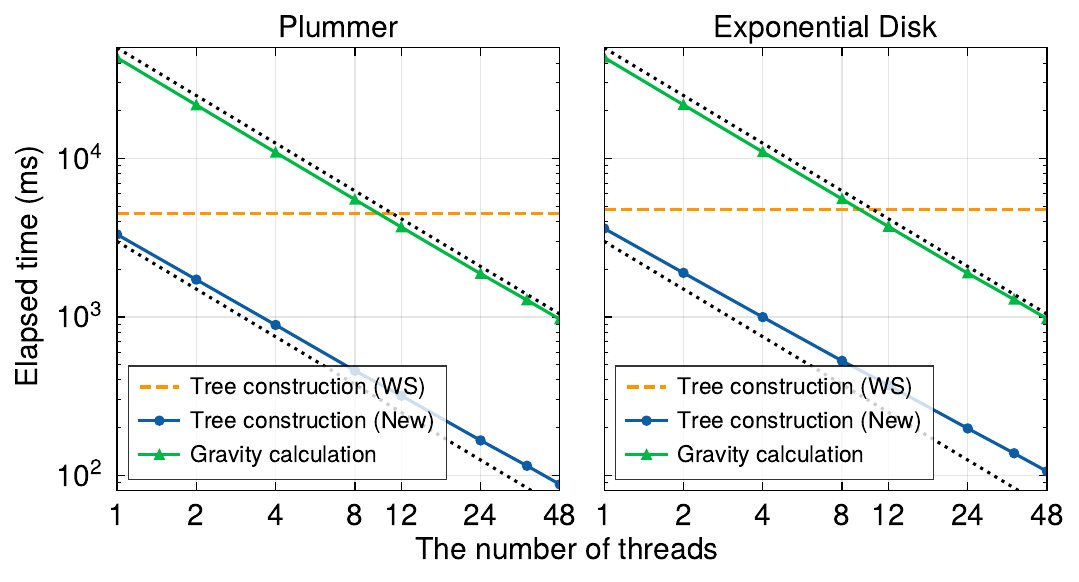}
        \caption{The Fugaku results.}
        \label{fig:fugaku_total_time}
    \end{subfigure}
    \begin{subfigure}{0.8\linewidth}
        \includegraphics[width=\linewidth]{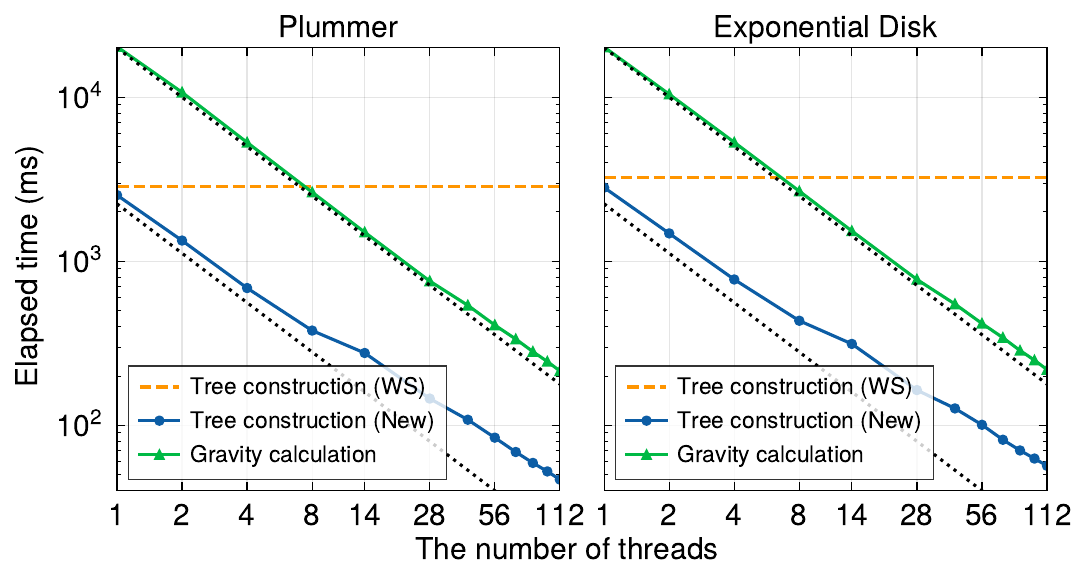}
        \caption{The Intel machine results.}
        \label{fig:prepost_total_time}
    \end{subfigure}
    \caption{Elapsed time for overall tree construction and gravity calculation of the tree algorithm. The figures contrast the time differences between our parallel tree construction method and the conventional sequential method. The number of particles is fixed at 16,777,216. The black dotted lines show ideal scalings.}
    \label{fig:parallel_total}
\end{figure*}

Then, we show the results of the performance tests of our code on multiple NUMA nodes. Fig.~\ref{fig:fugaku_parallel2} shows its performance on Fugaku across multiple CMGs, each containing 12 threads. Fig.~\ref{fig:fugaku_tree_time2} shows that the average elapsed time for all the tree construction components monotonically decreases as the number of CMGs increases, and the sorting of particle keys always takes the most time. Although there is no significant difference in the other three components, the reordering of particles takes the second most time in the exponential disk, while the creation of cells takes the second most time in the Plummer.

Fig.~\ref{fig:fugaku_tree_eff2} shows the parallel efficiency relative to one CMG. The initialization of the tree scales almost perfectly with the number of CMGs, indicating little overhead for parallelization and load imbalance. The decline of the reordering of particles is relatively mild compared to the results within one CMG shown in Fig.~\ref{fig:fugaku_tree_eff1}. It keeps about 80\,\% efficiency up to 4 CMGs. The total memory bandwidth increases with the number of CMGs, while being fixed within a CMG, causing this difference. The parallel efficiency of the sorting of particle keys and the creation of cells decreases slightly but remains above 85\,\%, and 87\,\%, respectively. The overall parallel efficiency of tree construction is over 90\,\% with the Plummer and above 87\,\% with the exponential disk.

Fig.~\ref{fig:prepost_parallel2} shows the results on the Intel machine across multiple NUMA nodes, each containing 14 threads. Unlike Fugaku, the number of CPUs increases by one with every two additional NUMA nodes. The results are considerably different from those on Fugaku. Fig.~\ref{fig:prepost_tree_time2} shows that the reordering of particles is the most time-consuming process, and the sorting of particle keys is the next, while it dominates on Fugaku. For four or more NUMA nodes, the time for creating cells is comparable to the time for sorting keys. Fig.~\ref{fig:prepost_tree_eff2} shows that the parallel efficiency of the initialization of the tree and the reordering of particles, which are embarrassingly parallel processes, does not drop significantly. When the number of NUMA nodes is small, the efficiency of the sorting of particle keys scales super-linearly. We attribute this to improved cache performance due to fewer particles and uniform memory performance per NUMA node. However, the efficiency of the sorting of particle keys decreases as the number of NUMA nodes increases, probably due to the tradeoff between the overhead for parallelization, including load imbalance, and the cache advantages. The parallel efficiency of the creation of cells declines steeply with fewer than four NUMA nodes and is asymptotically constant. Based on the fact that this decline occurs even with one CPU (two or fewer NUMA nodes) and becomes gradual with more NUMA nodes, it would not be due to the machine being composed of multiple CPUs but due to limited memory bandwidth relative to the number of particles. Although the low parallel efficiency of the creation of cells impacts the overall parallel efficiency to some extent, it still keeps above 69\,\% with up to 8 NUMA nodes.

Finally, we discuss the impact of our implementation on the overall performance of the tree algorithm (including both tree construction and gravity calculation). Fig.~\ref{fig:parallel_total} shows the average elapsed time per step for the tree algorithm divided into the tree construction and the gravity calculation. As expected, the gravity calculation dominates, and its parallel efficiency is outstandingly high, reaching approximately 93\,\% on 48 threads of Fugaku and exceeding 82\,\% on 112 threads of the Intel machine. On the other hand, the time taken for our parallel tree construction method shows deviations from the ideal scaling to some degree, but it still keeps steadily declining with increasing the number of threads. Assuming the conventional method is not multi-threaded, our novel method achieves 45 to 51 speedup on 48 threads of Fugaku and 56 to 61 speedup on 112 threads of the Intel machine over the conventional method. Unless the tree construction is highly parallelized, it is easy to see that it becomes a bottleneck from around ten threads regardless of the machines and the particle distributions, reinforcing the effectiveness of our implementation.

\section{Conclusions and Discussions}\label{sec:conclusions}
We have developed a novel tree construction method based on the hashed oct-tree proposed by~\citet{warren_parallel_1993}, which utilizes a different data structure from the original tree algorithm by \citet{barnes_hierarchical_1986}. Unlike the conventional method, our method assigns each particle to an appropriate leaf cell without traversing the tree and calculates the properties of cells as soon as they are created. We evaluated the performance of our method and compared it with the conventional method on two modern many-core processors.

For performance tests on a single thread, on the supercomputer Fugaku, we have achieved a speedup of at least 3.0 for the creation of cells, which was previously the most time-consuming part of the conventional method. As a result, the percentage of it in the total tree construction time is reduced to about 21\,\%. The time taken for creating cells with the conventional method on the Intel machine is about one-third of that on Fugaku, making it difficult to achieve significant speedup. Nonetheless, the speedup factor is above 2.2 on the Intel machine, and as a result, the time for creating cells occupies about 13\,\% of the total.

Then, we performed parallel performance tests on a single NUMA node. As the number of threads increases within a single NUMA node, the memory bandwidth per thread decreases, causing a drop in the parallel efficiency of memory-intensive procedures, particularly on the Intel machine, due to lower memory bandwidth. Nevertheless, the overall parallel efficiency of the tree construction keeps 81\,\% using 12 threads on Fugaku and is above 63\,\% using 14 threads on the Intel machine.

For parallel performance tests on multiple NUMA nodes, the decline in the parallel efficiency is more gradual. The overall efficiency maintains 86\,\% on 4 CMGs (48 threads) of Fugaku and 69\,\% on 8 NUMA nodes (112 threads) of the Intel machine. As a result, our method can speed up the tree construction by over 45 times on 48 threads of Fugaku and 56 times on 112 threads of the Intel machine compared to the conventional method. On the other hand, the conventional tree construction could become a bottleneck on such large threads as compared with the gravity calculation, which usually dominates the overall time of the tree algorithm. Those results reinforce the usefulness of our novel tree construction method. Our method should also be applicable to other particle simulations using tree structures, such as smoothed-particle hydrodynamics.

Our results indicate that the memory performance is crucial for the parallel tree construction. When we maintain the memory bandwidth per thread, increasing the number of threads does not significantly decrease the parallel efficiency. On Fugaku, regardless of the number of threads, the time for sorting keys dominates. In contrast, on the Intel machine, the high parallel efficiency of the sorting of particle keys emphasizes the importance of optimizing the reordering of particles and the creation of cells.

The number of CPU cores and SIMD width are expected to increase in future processors, confronting us with further optimizations of the tree construction. One approach to speed up the reordering of particles is to reduce its frequency. For instance, performing it once every ten steps will decrease the average elapsed time to about one-tenth. However, this optimization might increase cache misses when accessing particles. Another approach is the utilization of software prefetching. Since the destination indices of particles are stored in an array, issuing prefetch instructions at the right time might hide the data transfer latency.

\section*{Acknowledgements}
The authors wish to acknowledge the referee of this paper.
This work has been supported by IAAR Research Support Program in Chiba University Japan,
MEXT/JSPS KAKENHI (Grant Number JP21H01122 and JP23H04002),
MEXT as ``Program for Promoting Researches on the Supercomputer Fugaku''
(JPMXP1020200109 and JPMXP1020230406), and JICFuS.
We used computational resources of the supercomputer Fugaku provided by the RIKEN Center for Computational Science (Project ID: hp210164, hp220173, hp230173, and hp230204).

\section*{Data Availability}
The data and code underlying this article will be shared on reasonable request to the corresponding author.

\bibliographystyle{mnras}



\appendix

\section{Proof of the Leaf Depth Formula}\label{sec:proof_of_the_formula}
Particles $i$ and $j$ are in the same cell at any depth below $\text{minDepth}(i,j)$ and in different cells at or above this depth. Therefore, they are in different leaves if and only if the depth of the leaf containing particle $i$ is at least $\text{minDepth}(i,j)$. For particle $k$, which is the first particle in a leaf, particles 0 to $k-1$ are not in the same leaf. Thus, the leaf depth $D_k$ satisfies the following conditions:
\begin{equation}\label{eq:Dk_first_condition}
D_k\geq\text{minDepth}(k,0),\dots,D_k\geq\text{minDepth}(k,k-1).
\end{equation}
Since the keys of particles in the same cell are contiguous, if $i<j<k$, it cannot happen for particle $i$ to be in the same cell as particle $k$ while particle $j$ is not. In other words,
\begin{equation}
\text{minDepth}(k,0)\leq\cdots\leq\text{minDepth}(k,k-1).
\end{equation}
Thus, equation~\eqref{eq:Dk_first_condition} is equivalent to
\begin{equation}
D_k\geq\text{minDepth}(k,k-1).
\end{equation}
Furthermore, particle $k+n_\mathrm{leaf}$ and subsequent particles cannot be in the same cell as particle $k$ because the number of particles in any leaf is at most $n_\mathrm{leaf}$. Consequently,
\begin{equation}
D_k\geq\text{minDepth}(k,k+n_\mathrm{leaf}).
\end{equation}
In summary, the necessary conditions for the leaf to have at most $n_\mathrm{leaf}$ particles are
\begin{equation}\label{eq:Dk_final_condition}
D_k\geq\text{minDepth}(k,k-1), D_k\geq\text{minDepth}(k,k+n_\mathrm{leaf}).
\end{equation}
$D_k$ is the depth at which the number of particles first becomes $n_\mathrm{leaf}$ or fewer, and the parent cell at depth $D_k-1$ contains more than $n_\mathrm{leaf}$ particles by definition. Therefore, $D_k$ is the minimum depth that satisfies equation~\eqref{eq:Dk_final_condition}:
\begin{equation}
D_k=\max\{\text{minDepth}(k,k-1),\text{minDepth}(k,k+n_\mathrm{leaf})\}.
\end{equation}


\bsp	
\label{lastpage}
\end{document}